\documentclass[pra,letter,showpacs,twocolumn,nofootinbib]{revtex4}
\usepackage{graphicx}
\setkeys{Gin}{draft=false}
\usepackage{bm,amssymb,amsmath}
\usepackage{dcolumn}
\usepackage{natbib}
\usepackage{graphicx}

\newcommand{\fref}[1]{Fig.~\ref{#1}}

\newcommand{\tref}[1]{Table~\ref{#1}}
\newcommand{\rtw}{\longrightarrow}

\def\veps{\varepsilon}
\newcommand{\cm}{cm$^{-1}$}

\begin{document}

\title{Correlation, Breit, and QED effects in spectra of Mg-like ions}
\author{E. A. Konovalova$^{1}$}
\author{M. G. Kozlov$^{1,2}$}

\affiliation{$^1$Petersburg Nuclear Physics Institute, Gatchina 188300,
Russia}

\affiliation{$^2$St.~Petersburg Electrotechnical University ``LETI'', Prof.
Popov Str. 5, 197376 St.~Petersburg}

\date{
\today}

\begin{abstract}
We calculated spectra of the first six members of the Mg-like isoelectronic sequence using different approximations. The most accurate results were obtained with the configuration interaction + all order method (CI+AO), which provided relative accuracy for transition energies on the level of 0.1\%, or better. On this level of accuracy the Breit and QED corrections become important for the systems with nuclear charge $Z\gtrsim 20$. The retardation part of the Breit interaction and QED corrections for the partial waves with $l\ne 0$ are still too small to be important.

\end{abstract}

\maketitle
\section{Introduction}

High accuracy atomic calculations are necessary not only for atomic physics itself, but also for different applications from atomic frequency standards to tests of fundamental symmetries, search for the variation of the fundamental constants, and astrophysics. Atomic Mg and ions of the Mg isoelectronic sequence are convenient systems to test theoretical methods for atomic calculations \cite{DFK96b,JF97a,JSS97,JFG99,PKRD01,SJ02,ADF04,BDFK04,Koz04,BFK05,SKJJ09}. With ten core and two valence electrons they require accurate treatment of the core, core-valence, and valence correlations. Relativistic and quantum electrodynamic (QED) corrections are very small for magnesium, but grow along the isoelectronic series. Therefore in the high accuracy calculations of the heavier ions we need to account for both electronic correlations and relativistic and QED corrections.

At present there are several methods to calculate spectra and other properties of the many-electron atoms. The many-body perturbation theory (MBPT) is quite effective for monovalent atoms \cite{BJS90a,SSSJ03}. However, the order-by-order approach leads to rather slow convergence and becomes impractical above the third, or the fourth order. Because of that different variants of the all-order (AO) methods are currently used instead \cite{Koz03}. The coupled cluster (CC) method is the most popular all-order method, which is used by several groups \cite{BJS91,EKI94,MET00,CSD03,SCD06,DJ07,PBD10,KNSDV11,PatSah14}.

The MBPT-based approaches, including the CC method are most effective in treating core and core-valence correlations, but are less suitable for treating valence correlations, where there is no well defined small parameter. Here the non-perturbative approaches, such as configuration interaction (CI) \cite{KT87}, multi-configuration Hartree-Fock (MCHF), or multi-configuration Dirac-Hartree-Fock (MCDHF) \cite{JF97a,JFG99} are more effective. They allow accurate treatment of correlations between several valence electrons, but start to fail when the number of correlated electrons exceeds four, or five.

There are also mixed approaches where core and core-valence correlations are treated perturbatively and valence correlations are treated within CI method. The simplest method of this type includes CI and the second order MBPT (CI+MBPT) \cite{DFK96b,SJ02,Dzu05a}. The more advanced variant includes CI and linerized CC (CI+AO) \cite{Koz04,SKJJ09}. In both cases we use either MBPT, or CC  method to form an effective Hamiltonian for the valence electrons and then we use CI method to find valence energies and many-electron valence wave functions.

Relativistic effects for many-electron atoms are usually included within Dirac-Coulomb, or Dirac-Coulomb-Breit no-pair approximation. Sometimes, one-electron QED corrections (Lamb shifts) are also included using effective, or model potentials \cite{CCS00,SC11,RDF13,TuBe13,STY13}.

In this paper we study relative size of different corrections to the transition energies in Mg and Mg-like ions up to Cl VI. We use CI+MBPT and CI+AO methods for Dirac-Coulomb and Dirac-Coulomb-Breit Hamiltonians \cite{DFK96b,SKJJ09}.  Lamb shift corrections are included only for the $s$-electrons. We find that the relative accuracy of the CI+AO method for the transition frequencies is on the order of 0.1\% and remains almost constant along the sequence. At the same time, the dominant theoretical errors are different in the beginning and in the end of the sequence. While the role of the higher order correlation corrections is decreasing, the role of the relativistic and QED corrections grows.

\section{Theory and Method}

In this paper we use variants of the CI+MBPT and CI+AO methods, which are based on the Brillouin-Wigner perturbation theory. The resultant effective Hamiltonian for the valence electrons is symmetric, but energy-dependent \cite{DFK96b,SKJJ09}. \citet{SJ02} suggested an alternative approach based on the Rayleigh-Schr\"odinger perturbation theory. In this case the effective Hamiltonian is non-symmetric and does not depend on the energy.

We do not use variant of Savukov and Johnson because of the well-known problem of intruder states.
In our approach we do not have intruder states, but we need, in principle, to account for the energy-dependence of the effective Hamiltonian. Fortunately, if one is interested in the low-lying atomic states and forms effective Hamiltonian according to the recipe from Ref.\ \cite{DFK96b}, the respective corrections are very small. Thus, in the first approximation one can calculate MBPT diagrams at fixed energies. These energies are chosen to be the Hartree-Fock energies of the first valence orbitals with given symmetry (i.e. for Mg-like ions the energy $\veps_{3s}$ is used for all valence $s$ electrons, etc.) \cite{DFK96b}. In the next approximation one can also calculate first derivatives of all diagrams in respect to the energy. The effective Hamiltonian for a given energy is then formed by extrapolating diagrams to this energy \cite{DFK96b,KP99tr,Koz03}. Here we use this method to study respective corrections to the theoretical spectrum. We conclude, that at the present level of accuracy these corrections do not improve agreement with the experiment and can be neglected.

Our CI and CI+MBPT calculations are done with the package described in Ref.\ \cite{KPST15}. CI+AO calculations are done with the extended variant of the same package. These calculations include several steps. At first we solve linearized coupled cluster equations for the core and core-valence cluster amplitudes in the single-double (SD) approximation. Then we form one-electron and two-electron valence cluster amplitudes. On the third step we use these amplitudes to form effective valence Hamiltonian (see Ref.\ \cite{SKJJ09} for details). In our opinion the CI+AO method described above treats valence-valence, valence-core, and core-core correlations in a most effective manner.

\subsection*{Calculation details}

We calculate spectrum of each ion of the isoelectronic sequence within several approximations, namely pure valence CI, CI+MBPT, and CI+AO methods. We use Dirac-Coulomb Hamiltonian in the no-pair approximation. We also do calculations with Breit and QED corrections. Finally, for the CI+MBPT and CI+AO we repeat computations with account for the energy dependence of the effective Hamiltonian.  Previously this dependence has been studied for the CI+MBPT \cite{DFK96b,KP99tr}, but not for the CI+AO method. At the same time, the consistent treatment of the high orders requires accurate account of the energy dependence of the effective Hamiltonian \cite{Koz03}.

For each ion we form the finite basis set, which includes Dirac-Fock orbitals for the core and valence states and Sturmian orbitals for virtual states. Sturmian orbitals effectively account for both discrete and continuous spectrum \cite{TL03}. After adding Sturmian orbitals we diagonalize Dirac-Fock operator of the core on the whole basis set (i.e. we use the $V^{N-2}$ potential, where $N=12$ is the total number of electrons).

This way we make $30spdfgh$ basis set of 300 orbitals that includes partial waves with $l=0,\ldots,5$. We use the whole basis set for the MBPT and CC parts of the calculation. The configuration space for the CI part includes single (S) and double (D) excitations to all orbitals up to $21spdfg$. This corresponds to the full two-electron CI on the basis set $21spdfg$. As no core excitations are included in the CI space, in the pure CI calculation we neglect all core-valence and core-core correlations. In the CI+MBPT and CI+AO calculations these correlations are included in the effective Hamiltonian of the valence electrons.

\begin{table*}[tH]
\caption{
Calculated spectra of Mg I and Al II. Experimental data are from
Ref.\ \cite{NIST}. Theoretical results are presented in columns 4 -- 7: CI is the
calculation within configuration interaction method, MBPT -- combined CI and second order MBPT method, and AO~--~combined CI and all-order method. Br+QED stands for contribution of the Breit interaction and QED. For the ground state we give binding energy of the two valence electrons (the sum of the first two ionization potentials). For other states we give excitation energies from the ground state. All energies are in \cm. Next three columns give relative errors in percent for different methods. The Final theoretical values (CI+all-order+Breit+QED) and their errors are listed in the last three columns. For the final error we give both relative (in \%) and absolute value (in \cm).
}

\label{tbl1:energies}
\begin{tabular}{clrrrrrrrrrrr@/r}
\hline \hline
\multicolumn{1}{c}{ Element} & \multicolumn{2}{c}{State} & \multicolumn{1}{c}{Expt.}  & \multicolumn{1}{c}{CI} & \multicolumn{1}{c}{MBPT} & \multicolumn{1}{c}{AO}  &  \multicolumn{1}{c}{Br+QED} & \multicolumn{3}{c} {Diff. with expt.(\%)} & \multicolumn{3}{c} {Final theory} \\
&    &   &    &   &  &  &  \multicolumn{1}{c}{contrib.} & \multicolumn{1}{c}{CI} & \multicolumn{1}{c}{MBPT} & \multicolumn{1}{c}{AO} &  \multicolumn{3}{c}{AO+Br+QED}\\
\hline
 Mg I & $3s{^2}$ & ${^1}S_0$ &  182939 & 179554  & 182685  & 182875  &  $-$47 \quad \quad & 1.9 & 0.14 &     0.035 & \quad 182828 &  0.060\,\, & 111 \\
        & $3s3p$ & ${^3}P_0$ &   21850 &   20919 &   21792 &   21851 &  $-$10 \quad \quad & 4.3 & 0.27 &  $-$0.002 & \quad  21841 &  0.042\,\, & 9   \\
        & $3s3p$ & ${^3}P_1$ &   21871 &   20939 &   21814 &   21872 &  $-$11 \quad \quad & 4.3 & 0.26 &  $-$0.009 & \quad  21862 &  0.040\,\, & 9   \\
        & $3s3p$ & ${^3}P_2$ &   21911 &   20980 &   21857 &   21915 &  $-$13 \quad \quad & 4.2 & 0.25 &  $-$0.021 & \quad  21903 &  0.037\,\, & 8   \\
        & $3s3p$ & ${^1}P_1$ &   35051 &   34471 &   35030 &   35052 &  $-$15 \quad \quad & 1.7 & 0.06 &  $-$0.002 & \quad  35037 &  0.040\,\, & 14  \\
        & $3s4s$ & ${^3}S_1$ &   41197 &   40419 &   41132 &   41184 &  $-$14 \quad \quad & 1.9 & 0.16 &     0.032 & \quad  41170 &  0.066\,\, & 27  \\
        & $3s4s$ & ${^1}S_0$ &   43503 &   42678 &   43438 &   43490 &  $-$14 \quad \quad & 1.9 & 0.15 &     0.031 & \quad  43475 &  0.064\,\, & 28  \\
        & $3s3d$ & ${^1}D_2$ &   46403 &   45126 &   46288 &   46364 &  $-$20 \quad \quad & 2.8 & 0.25 &     0.085 & \quad  46344 &  0.127\,\, & 59  \\
        & $3s4p$ & ${^3}P_0$ &   47841 &   46932 &   47766 &   47822 &  $-$16 \quad \quad & 1.9 & 0.16 &     0.041 & \quad  47806 &  0.073\,\, & 35  \\
        & $3s4p$ & ${^3}P_1$ &   47844 &   46935 &   47769 &   47825 &  $-$16 \quad \quad & 1.9 & 0.16 &     0.040 & \quad  47809 &  0.073\,\, & 35  \\
        & $3s4p$ & ${^3}P_2$ &   47851 &   46942 &   47776 &   47832 &  $-$16 \quad \quad & 1.9 & 0.16 &     0.039 & \quad  47816 &  0.073\,\, & 35  \\
        & $3s3d$ & ${^3}D_2$ &   47957 &   46972 &   47851 &   47904 &  $-$18 \quad \quad & 2.1 & 0.22 &     0.110 & \quad  47886 &  0.147\,\, & 71  \\
        & $3s3d$ & ${^3}D_3$ &   47957 &   46972 &   47851 &   47904 &  $-$18 \quad \quad & 2.1 & 0.22 &     0.110 & \quad  47886 &  0.147\,\, & 71  \\
        & $3s3d$ & ${^3}D_1$ &   47957 &   46972 &   47851 &   47904 &  $-$18 \quad \quad & 2.1 & 0.22 &     0.110 & \quad  47886 &  0.147\,\, & 71  \\
        & $3s4p$ & ${^1}P_1$ &   49347 &   48497 &   49284 &   49332 &  $-$16 \quad \quad & 1.7 & 0.13 &     0.030 & \quad  49316 &  0.063\,\, & 31  \\
[1.5mm]

Al II & $3s{^2}$ & ${^1}S_0$ &  381308 &  376617 &  381166 & 381313 & $-$103 \quad \quad & 1.2 &   0.04 &   $-$0.001 & \quad 381210  &     0.026\,\, &    98  \\
        & $3s3p$ & ${^3}P_0$ &   37393 &   36257 &   37354 &  37406 &  $-$14 \quad \quad & 3.0 &   0.10 &   $-$0.035 & \quad  37392  &     0.003\,\, &    1  \\
        & $3s3p$ & ${^3}P_1$ &   37454 &   36319 &   37419 &  37471 &  $-$18 \quad \quad & 3.0 &   0.09 &   $-$0.046 & \quad  37454  &     0.000\,\, &    0  \\
        & $3s3p$ & ${^3}P_2$ &   37578 &   36444 &   37550 &  37603 &  $-$24 \quad \quad & 3.0 &   0.07 &   $-$0.066 & \quad  37579  &  $-$0.003\,\, & $-$1  \\
        & $3s3p$ & ${^1}P_1$ &   59852 &   59538 &   59900 &  59882 &  $-$27 \quad \quad & 0.5 &$-$0.08 &   $-$0.050 & \quad  59855  &  $-$0.005\,\, & $-$3  \\
      & $3p{^2}$ & ${^1}D_2$ &   85481 &   83516 &   85415 &  85494 &  $-$45 \quad \quad & 2.3 &   0.08 &   $-$0.015 & \quad  85450  &     0.037\,\, &    32 \\
        & $3s4s$ & ${^3}S_1$ &   91275 &   90010 &   91238 &  91288 &  $-$32 \quad \quad & 1.4 &   0.04 &   $-$0.015 & \quad  91256  &     0.020\,\, &    19 \\
      & $3p{^2}$ & ${^3}P_0$ &   94085 &   92598 &   94060 &  94087 &  $-$38 \quad \quad & 1.6 &   0.03 &   $-$0.003 & \quad  94049  &     0.038\,\, &    36 \\
      & $3p{^2}$ & ${^3}P_1$ &   94147 &   92661 &   94126 &  94153 &  $-$42 \quad \quad & 1.6 &   0.02 &   $-$0.006 & \quad  94112  &     0.038\,\, &    36 \\
      & $3p{^2}$ & ${^3}P_2$ &   94269 &   92783 &   94255 &  94283 &  $-$48 \quad \quad & 1.6 &   0.01 &   $-$0.015 & \quad  94234  &     0.036\,\, &    34 \\
        & $3s4s$ & ${^1}S_0$ &   95351 &   94012 &   95320 &  95369 &  $-$33 \quad \quad & 1.4 &   0.03 &   $-$0.020 & \quad  95336  &     0.015\,\, &    15 \\
        & $3s3d$ & ${^3}D_3$ &   95549 &   94171 &   95443 &  95463 &  $-$45 \quad \quad & 1.4 &   0.11 &      0.091 & \quad  95418  &     0.138\,\, &    131 \\
        & $3s3d$ & ${^3}D_2$ &   95551 &   94172 &   95444 &  95464 &  $-$45 \quad \quad & 1.4 &   0.11 &      0.091 & \quad  95419  &     0.138\,\, &    131 \\
        & $3s3d$ & ${^3}D_1$ &   95551 &   94172 &   95444 &  95464 &  $-$44 \quad \quad & 1.4 &   0.11 &      0.091 & \quad  95420  &     0.138\,\, &    132 \\
        & $3s4p$ & ${^3}P_0$ &  105428 &  103965 &  105379 & 105431 &  $-$35 \quad \quad & 1.4 &   0.05 &   $-$0.004 & \quad 105396  &     0.030\,\, &    32 \\
        & $3s4p$ & ${^3}P_1$ &  105442 &  103979 &  105393 & 105446 &  $-$36 \quad \quad & 1.4 &   0.05 &   $-$0.004 & \quad 105410  &     0.030\,\, &    31 \\
        & $3s4p$ & ${^3}P_2$ &  105471 &  104009 &  105424 & 105477 &  $-$38 \quad \quad & 1.4 &   0.04 &   $-$0.006 & \quad 105440  &     0.030\,\, &    31 \\
        & $3s4p$ & ${^1}P_1$ &  106921 &  105578 &  106885 & 106929 &  $-$37 \quad \quad & 1.3 &   0.03 &   $-$0.008 & \quad 106892  &     0.027\,\, &    28 \\

\hline \hline
\end{tabular}
\end{table*}

\begin{table*}[tH]
\caption{Same as in Table \ref{tbl1:energies}  for Si III and P IV .}

\label{tbl2:energies}
\begin{tabular}{clrrrrrrrrrrr@/rl}
\hline \hline
\multicolumn{1}{c}{ Element} & \multicolumn{2}{c}{State} & \multicolumn{1}{c}{Expt.}  & \multicolumn{1}{c}{CI} & \multicolumn{1}{c}{MBPT} & \multicolumn{1}{c}{AO}  &  \multicolumn{1}{c}{Br+QED} & \multicolumn{3}{c} {Diff. with expt.(\%)} & \multicolumn{3}{c} {Final theory} \\
&    &   &    &   &  &  &  \multicolumn{1}{c}{contrib.} & \multicolumn{1}{c}{CI} & \multicolumn{1}{c}{MBPT} & \multicolumn{1}{c}{AO} &  \multicolumn{3}{c}{AO+Br+QED}\\

\hline

Si III& $3s{^2}$ & ${^1}S_0$ &  634232 &  628587 &  634174 &  634298 &  $-$181\quad \quad &   0.9 &   0.01 & $-$0.010 & \quad 634117  &    0.018\,\, &     115 \\
        & $3s3p$ & ${^3}P_0$ &   52725 &   51546 &   52702 &   52750 &  $-$18 \quad \quad &   2.2 &   0.04 & $-$0.049 & \quad  52733  & $-$0.015\,\, &  $-$8  \\
        & $3s3p$ & ${^3}P_1$ &   52853 &   51678 &   52838 &   52887 &  $-$24 \quad \quad &   2.2 &   0.03 & $-$0.063 & \quad  52863  & $-$0.018\,\, &  $-$9  \\
        & $3s3p$ & ${^3}P_2$ &   53115 &   51944 &   53114 &   53163 &  $-$37 \quad \quad &   2.2 &   0.00 & $-$0.090 & \quad  53126  & $-$0.021\,\, &  $-$11 \\
        & $3s3p$ & ${^1}P_1$ &   82884 &   83001 &   82978 &   82942 &  $-$39 \quad \quad &$-$0.1 &$-$0.11 & $-$0.069 & \quad  82903  & $-$0.023\,\, &  $-$19 \\
      & $3p{^2}$ & ${^1}D_2$ &  122215 &  120238 &  122216 &  122286 &  $-$69 \quad \quad &   1.6 &   0.00 & $-$0.059 & \quad 122217  & $-$0.002\,\, &  $-$2  \\
      & $3p{^2}$ & ${^3}P_0$ &  129708 &  128580 &  129724 &  129731 &  $-$52 \quad \quad &   0.9 &$-$0.01 & $-$0.018 & \quad 129679  &    0.022\,\, &     29 \\
      & $3p{^2}$ & ${^3}P_1$ &  129842 &  128715 &  129864 &  129872 &  $-$59 \quad \quad &   0.9 &$-$0.02 & $-$0.023 & \quad 129813  &    0.022\,\, &     29 \\
      & $3p{^2}$ & ${^3}P_2$ &  130101 &  128977 &  130138 &  130146 &  $-$72 \quad \quad &   0.9 &$-$0.03 & $-$0.035 & \quad 130075  &    0.020\,\, &     26 \\
        & $3s3d$ & ${^3}D_3$ &  142944 &  141714 &  142867 &  142854 &  $-$81 \quad \quad &   0.9 &   0.05 &    0.063 & \quad 142773  &    0.119\,\, &     170 \\
        & $3s3d$ & ${^3}D_2$ &  142946 &  141715 &  142867 &  142854 &  $-$79 \quad \quad &   0.9 &   0.05 &    0.064 & \quad 142775  &    0.119\,\, &     171 \\
        & $3s3d$ & ${^3}D_1$ &  142948 &  141716 &  142868 &  142855 &  $-$78 \quad \quad &   0.9 &   0.06 &    0.065 & \quad 142777  &    0.120\,\, &     172 \\
        & $3s4s$ & ${^3}S_1$ &  153377 &  151780 &  153378 &  153425 &  $-$57 \quad \quad &   1.0 &   0.00 & $-$0.031 & \quad 153368  &    0.006\,\, &     9  \\
      & $3p{^2}$ & ${^1}S_0$ &  153444 &  152676 &  153628 &  153610 &  $-$70 \quad \quad &   0.5 &$-$0.12 & $-$0.108 & \quad 153540  & $-$0.062\,\, &  $-$96  \\
        & $3s4s$ & ${^1}S_0$ &  159070 &  157563 &  159096 &  159136 &  $-$60 \quad \quad &   0.9 &$-$0.02 & $-$0.042 & \quad 159076  & $-$0.004\,\, &  $-$7  \\
        & $3s3d$ & ${^1}D_2$ &  165765 &  165093 &  165825 &  165778 &  $-$75 \quad \quad &   0.4 &$-$0.04 & $-$0.008 & \quad 165703  &    0.037\,\, &    62 \\
        & $3s4p$ & ${^3}P_0$ &  175230 &  173427 &  175215 &  175264 &  $-$61 \quad \quad &   1.0 &   0.01 & $-$0.019 & \quad 175203  &    0.016\,\, &    27 \\
        & $3s4p$ & ${^3}P_1$ &  175263 &  173461 &  175250 &  175299 &  $-$63 \quad \quad &   1.0 &   0.01 & $-$0.021 & \quad 175236  &    0.015\,\, &    27 \\
        & $3s4p$ & ${^3}P_2$ &  175336 &  173536 &  175327 &  175376 &  $-$66 \quad \quad &   1.0 &   0.01 & $-$0.023 & \quad 175310  &    0.015\,\, &    26 \\
        & $3s4p$ & ${^1}P_1$ &  176487 &  174829 &  176487 &  176530 &  $-$65 \quad \quad &   0.9 &   0.00 & $-$0.024 & \quad 176465  &    0.013\,\, &    22 \\
[1.5mm]

P  IV & $3s{^2}$ & ${^1}S_0$ &  939386 &  933026 &  939426 &  939530 &  $-$283 \quad \quad &   0.7 &   0.00 & $-$0.015 & \quad 939247 &    0.015\,\, &    139\\
        & $3s3p$ & ${^3}P_0$ &   67918 &   66765 &   67907 &   67951 &  $-$20  \quad \quad &   1.7 &   0.02 & $-$0.048 & \quad  67931 & $-$0.019\,\, & $-$13 \\
        & $3s3p$ & ${^3}P_1$ &   68146 &   67001 &   68149 &   68192 &  $-$30  \quad \quad &   1.7 &   0.00 & $-$0.068 & \quad  68162 & $-$0.023\,\, & $-$16 \\
        & $3s3p$ & ${^3}P_2$ &   68615 &   67479 &   68641 &   68686 &  $-$52  \quad \quad &   1.7 &$-$0.04 & $-$0.103 & \quad  68634 & $-$0.027\,\, & $-$19 \\
        & $3s3p$ & ${^1}P_1$ &  105190 &  105754 &  105320 &  105277 &  $-$52  \quad \quad &$-$0.5 &$-$0.12 & $-$0.082 & \quad 105225 & $-$0.033\,\, & $-$34 \\
      & $3p{^2}$ & ${^1}D_2$ &  158142 &  156361 &  158207 &  158266 &  $-$96  \quad \quad &   1.1 &$-$0.04 & $-$0.079 & \quad 158170 & $-$0.018\,\, & $-$28 \\
      & $3p{^2}$ & ${^3}P_0$ &  164941 &  164237 &  164987 &  164984 &  $-$66  \quad \quad &   0.4 &$-$0.03 & $-$0.026 & \quad 164918 &    0.014\,\, &    23 \\
      & $3p{^2}$ & ${^3}P_1$ &  165185 &  164484 &  165243 &  165240 &  $-$77  \quad \quad &   0.4 &$-$0.04 & $-$0.033 & \quad 165163 &    0.014\,\, &    23 \\
      & $3p{^2}$ & ${^3}P_2$ &  165654 &  164962 &  165739 &  165736 &  $-$100 \quad \quad &   0.4 &$-$0.05 & $-$0.049 & \quad 165636 &    0.011\,\, &    18\\
        & $3s3d$ & ${^3}D_2$ &  189398 &  188560 &  189368 &  189335 &  $-$125 \quad \quad &   0.4 &   0.02 &    0.033 & \quad 189210 &    0.099\,\, &    188\\
        & $3s3d$ & ${^3}D_1$ &  189398 &  188557 &  189363 &  189330 &  $-$121 \quad \quad &   0.4 &   0.02 &    0.036 & \quad 189209 &    0.100\,\, &    189\\
        & $3s3d$ & ${^3}D_3$ &  189398 &  188555 &  189360 &  189327 &  $-$119 \quad \quad &   0.4 &   0.02 &    0.037 & \quad 189209 &    0.100\,\, &    189\\
      & $3p{^2}$ & ${^1}S_0$ &  194592 &  194530 &  194857 &  194821 &  $-$96  \quad \quad &   0.0 &$-$0.14 & $-$0.118 & \quad 194725 & $-$0.068\,\, & $-$133\\
        & $3s3d$ & ${^1}D_2$ &  219154 &  219299 &  219300 &  219215 &  $-$116 \quad \quad &$-$0.1 &$-$0.07 & $-$0.028 & \quad 219099 &    0.025\,\, &    55 \\
        & $3s4s$ & ${^3}S_1$ &  226898 &  225066 &  226947 &  226990 &  $-$89  \quad \quad &   0.8 &$-$0.02 & $-$0.040 & \quad 226901 & $-$0.001\,\, & $-$3  \\
        & $3s4s$ & ${^1}S_0$ &  233998 &  232162 &  234071 &  234110 &  $-$92  \quad \quad &   0.8 &$-$0.03 & $-$0.048 & \quad 234018 & $-$0.009\,\, & $-$20 \\
        & $3s4p$ & ${^3}P_0$ &  256553 &  254539 &  256585 &  256629 &  $-$94  \quad \quad &   0.8 &$-$0.01 & $-$0.029 & \quad 256535 &    0.007\,\, &    19 \\
        & $3s4p$ & ${^3}P_1$ &  256612 &  254602 &  256647 &  256691 &  $-$96  \quad \quad &   0.8 &$-$0.01 & $-$0.031 & \quad 256594 &    0.007\,\, &    18 \\
        & $3s4p$ & ${^3}P_2$ &  256760 &  254753 &  256803 &  256847 &  $-$104 \quad \quad &   0.8 &$-$0.02 & $-$0.034 & \quad 256743 &    0.007\,\, &    17 \\
        & $3s4p$ & ${^1}P_1$ &  257523 &  255651 &  257570 &  257610 &  $-$102 \quad \quad &   0.7 &$-$0.02 & $-$0.034 & \quad 257508 &    0.006\,\, & 14\\

\hline \hline
\end{tabular}
\end{table*}

\subsection*{QED corrections}

We use semiempirical approach to account for QED corrections where we include only the Lamb shift for the $s$-electrons
[It is known that Lamb shift for other partial waves is at least one order of magnitude smaller]. Following \cite{STY13} we can parametrize Lamb shift for the hydrogen-like ion as (we use atomic units $\hbar=e=m_e=1$):
 \begin{align}\label{LS1}
 \Delta \veps_\mathrm{QED} &= \frac{\alpha^3 Z^4}{\pi n^3}
 F_n(\alpha Z)\,,
 \end{align}
where $n$ is the principle quantum number, $\alpha$ is the fine structure constant, and $Z$ is the nuclear charge. This expression can be generalized for the non-diagonal matrix elements:
 \begin{align}\label{LS2}
 \sigma_{n,n'} &= \frac{\alpha^3 Z^4}{\pi (n\,n')^{3/2}}
 F_{n,n'}(\alpha Z)\,.
 \end{align}
Function $F$ weakly depends on $\alpha Z$ and on indexes $n,n'$. This function is tabulated in Ref.\ \cite{STY13}. Noting that $\frac{Z^3}{\pi (n\,n')^{3/2}}=|\Psi_n(0)\Psi_{n'}(0)|$ we get:
 \begin{align}\label{LS3}
 \sigma_{n,n'} &= \frac{\alpha^3 Z}{\pi}\left|\Psi_n(0)\Psi_{n'}(0)\right|
 F_{n,n'}(\alpha Z)\,.
 \end{align}
This expression can be used not only for hydrogen-like ions, but also for the valence electrons of many-electron atoms. Following Ref.\ \cite{Sob79} we can express electron density at the origin in terms of the binding energy $\veps_\nu$:
 \begin{align}
 \veps_\nu = \tfrac{(Z_i+1)^2}{2 \nu^2}\,; &
\nonumber \\
\left|\Psi_\nu(0)\Psi_{\nu'}(0)\right| &=
 \frac{(Z_i +1)^2 Z}{\pi (\nu \nu')^{3/2}}
\nonumber \\
\label{LS4}
&=
\frac{(4 \veps_\nu \veps_{\nu'})^{3/4} Z}{\pi (Z_i +1)}
\,.
 \end{align}
In these expressions $Z_i$ is the charge of the ion (for neutral atom $Z_i=0$) and $\nu$ is the effective quantum number. Now we can write QED corrections as:
 \begin{align}\label{LS6}
 \sigma_{\nu,\nu'} &= \frac{(4 \veps_\nu \veps_{\nu'})^{3/4}}{\pi (Z_i +1)}\, \alpha^3 Z^2 F(\alpha Z)\,.
 \end{align}
Binding energy $ \veps_\nu$ for the valence electrons in the many-electron atoms and ions corresponds to the very large quantum numbers $n$ of the hydrogen-like ions with the same $Z$. Therefore, the function $F(\alpha Z)$ in \eqref{LS6} corresponds to the limit $n,n'\rtw\infty$ for the function $F_{n,n'}(\alpha Z)$. According to \cite{STY13} this function slowly decreases with $Z$, but for the range $12 \le Z \le 17$ we can take $F(\alpha Z)\approx 4$. Thus, we can write:
 \begin{align}\label{LS7}
 \sigma_{\nu,\nu'} &\approx 8\sqrt{2}\frac{(\veps_\nu \veps_{\nu'})^{3/4}}{\pi (Z_i +1)}\, \alpha^3 Z^2 \,.
 \end{align}
In our CI+AO calculations we add these QED corrections to the one-electron radial integrals of valence electrons.

\begin{table*}[tH]
\caption{Same as in Table \ref{tbl1:energies}  for S V and Cl VI .}

\label{tbl3:energies}
\begin{tabular}{clrrrrrrrrrrr@/rl}
\hline \hline
\multicolumn{1}{c}{ Element} & \multicolumn{2}{c}{State} & \multicolumn{1}{c}{Expt.}  & \multicolumn{1}{c}{CI} & \multicolumn{1}{c}{MBPT} & \multicolumn{1}{c}{AO}  &  \multicolumn{1}{c}{Br+QED} & \multicolumn{3}{c} {Diff. with expt.(\%)} & \multicolumn{3}{c} {Final theory} \\
&    &   &    &   &  &  &  \multicolumn{1}{c}{contrib.} & \multicolumn{1}{c}{CI} & \multicolumn{1}{c}{MBPT} & \multicolumn{1}{c}{AO} &  \multicolumn{3}{c}{AO+Br+QED}\\

\hline

S  V  & $3s{^2}$ & ${^1}S_0$ & 1295709 & 1288803 & 1295858 & 1295948 &  $-$415 \quad \quad &   0.5 &$-$0.01 & $-$0.018 & \quad1295532 &   0.014\,\, &    176 \\
        & $3s3p$ & ${^3}P_0$ &   83024 &   81926 &   83020 &   83060 &  $-$21  \quad \quad &   1.3 &   0.00 & $-$0.044 & \quad  83039 &$-$0.018\,\, & $-$15 \\
        & $3s3p$ & ${^3}P_1$ &   83394 &   82307 &   83409 &   83449 &  $-$37  \quad \quad &   1.3 &$-$0.02 & $-$0.067 & \quad  83412 &$-$0.022\,\, & $-$18 \\
        & $3s3p$ & ${^3}P_2$ &   84155 &   83087 &   84208 &   84248 &  $-$70  \quad \quad &   1.3 &$-$0.06 & $-$0.111 & \quad  84178 &$-$0.027\,\, & $-$23 \\
        & $3s3p$ & ${^1}P_1$ &  127151 &  128134 &  127312 &  127265 &  $-$67  \quad \quad &$-$0.8 &$-$0.13 & $-$0.090 & \quad 127198 &$-$0.037\,\, & $-$48 \\
      & $3p{^2}$ & ${^1}D_2$ &  193739 &  192223 &  193859 &  193910 &  $-$126 \quad \quad &   0.8 &$-$0.06 & $-$0.088 & \quad 193784 &$-$0.023\,\, & $-$45 \\
      & $3p{^2}$ & ${^3}P_0$ &  199967 &  199690 &  200036 &  200027 &  $-$80  \quad \quad &   0.1 &$-$0.03 & $-$0.030 & \quad 199947 &   0.010\,\, &    20 \\
      & $3p{^2}$ & ${^3}P_1$ &  200371 &  200101 &  200460 &  200451 &  $-$99  \quad \quad &   0.1 &$-$0.04 & $-$0.040 & \quad 200352 &   0.009\,\, &    18 \\
      & $3p{^2}$ & ${^3}P_2$ &  201146 &  200891 &  201277 &  201268 &  $-$134 \quad \quad &   0.1 &$-$0.07 & $-$0.061 & \quad 201135 &   0.006\,\, &    11 \\
        & $3s3d$ & ${^3}D_1$ &  234942 &  234556 &  234947 &  234905 &  $-$166 \quad \quad &   0.2 &   0.00 &    0.016 & \quad 234739 &   0.086\,\, &    203 \\
        & $3s3d$ & ${^3}D_2$ &  234947 &  234566 &  234959 &  234917 &  $-$172 \quad \quad &   0.2 &$-$0.01 &    0.013 & \quad 234746 &   0.086\,\, &    201 \\
        & $3s3d$ & ${^3}D_3$ &  234956 &  234580 &  234977 &  234936 &  $-$180 \quad \quad &   0.2 &$-$0.01 &    0.009 & \quad 234756 &   0.085\,\, &    200 \\
      & $3p{^2}$ & ${^1}S_0$ &  235350 &  235873 &  235675 &  235633 &  $-$126 \quad \quad &$-$0.2 &$-$0.14 & $-$0.120 & \quad 235507 &$-$0.067\,\, & $-$157 \\
        & $3s3d$ & ${^1}D_2$ &  270700 &  271683 &  270933 &  270829 &  $-$165 \quad \quad &$-$0.4 &$-$0.09 & $-$0.048 & \quad 270664 &   0.013\,\, &    36 \\
        & $3s4s$ & ${^3}S_1$ &  311595 &  309588 &  311692 &  311730 &  $-$129 \quad \quad &   0.6 &$-$0.03 & $-$0.043 & \quad 311601 &$-$0.002\,\, & $-$6 \\
        & $3s4s$ & ${^1}S_0$ &  320108 &  318110 &  320238 &  320274 &  $-$134 \quad \quad &   0.6 &$-$0.04 & $-$0.052 & \quad 320140 &$-$0.010\,\, & $-$32 \\
        & $3p3d$ & ${^3}F_2$ &  323133 &  321355 &  323119 &  323130 &  $-$200 \quad \quad &   0.6 &   0.00 &    0.001 & \quad 322931 &   0.062\,\, &    202 \\
        & $3p3d$ & ${^3}F_3$ &  323547 &  321780 &  323559 &  323571 &  $-$224 \quad \quad &   0.5 &   0.00 & $-$0.007 & \quad 323348 &   0.062\,\, &    200 \\
        & $3p3d$ & ${^3}F_4$ &  324080 &  322331 &  324126 &  324139 &  $-$254 \quad \quad &   0.5 &$-$0.01 & $-$0.018 & \quad 323885 &   0.060\,\, &    195 \\
        & $3p3d$ & ${^1}D_2$ &  328454 &  327374 &  328614 &  328591 &  $-$234 \quad \quad &   0.3 &$-$0.05 & $-$0.042 & \quad 328357 &   0.030\,\, &    97 \\
        & $3p3d$ & ${^3}P_2$ &  345338 &  344679 &  345437 &  345389 &  $-$225 \quad \quad &   0.2 &$-$0.03 & $-$0.015 & \quad 345165 &   0.050\,\, &    173 \\
        & $3p3d$ & ${^3}P_1$ &  345713 &  345038 &  345818 &  345771 &  $-$232 \quad \quad &   0.2 &$-$0.03 & $-$0.017 & \quad 345539 &   0.050\,\, &    174 \\
        & $3p3d$ & ${^3}P_0$ &  345953 &  345255 &  346062 &  346016 &  $-$241 \quad \quad &   0.2 &$-$0.03 & $-$0.018 & \quad 345775 &   0.051\,\, &    178 \\
[1.5mm]

Cl IV & $3s{^2}$ & ${^1}S_0$ & 1702996 & 1695325 & 1702922 & 1703001 &  $-$580 \quad \quad &   0.5 &   0.00 &    0.000 & \quad1702421 &    0.034\,\, &    575\\
        & $3s3p$ & ${^3}P_0$ &   98062 &   97047 &   98079 &   98115 &  $-$22  \quad \quad &   1.0 &$-$0.02 & $-$0.054 & \quad  98093 & $-$0.032\,\, & $-$31 \\
        & $3s3p$ & ${^3}P_1$ &   98621 &   97623 &   98664 &   98701 &  $-$44  \quad \quad &   1.0 &$-$0.04 & $-$0.082 & \quad  98657 & $-$0.036\,\, & $-$36 \\
        & $3s3p$ & ${^3}P_2$ &   99782 &   98813 &   99879 &   99916 &  $-$92  \quad \quad &   1.0 &$-$0.10 & $-$0.134 & \quad  99824 & $-$0.042\,\, & $-$42 \\
        & $3s3p$ & ${^1}P_1$ &  148947 &  150326 &  149149 &  149102 &  $-$85  \quad \quad &$-$0.9 &$-$0.14 & $-$0.104 & \quad 149018 & $-$0.047\,\, & $-$71 \\
      & $3p{^2}$ & ${^1}D_2$ &  229219 &  228003 &  229404 &  229448 &  $-$160 \quad \quad &   0.5 &$-$0.08 & $-$0.100 & \quad 229288 & $-$0.030\,\, & $-$69 \\
      & $3p{^2}$ & ${^3}P_0$ &  234886 &  235038 &  234996 &  234983 &  $-$94  \quad \quad &$-$0.1 &$-$0.05 & $-$0.041 & \quad 234888 & $-$0.001\,\, & $-$2 \\
      & $3p{^2}$ & ${^3}P_1$ &  235518 &  235681 &  235655 &  235643 &  $-$123 \quad \quad &$-$0.1 &$-$0.06 & $-$0.053 & \quad 235520 & $-$0.001\,\, & $-$2 \\
      & $3p{^2}$ & ${^3}P_2$ &  236721 &  236907 &  236922 &  236911 &  $-$175 \quad \quad &$-$0.1 &$-$0.08 & $-$0.080 & \quad 236735 & $-$0.006\,\, & $-$14 \\
      & $3p{^2}$ & ${^1}S_0$ &  275988 &  277074 &  276389 &  276345 &  $-$162 \quad \quad &$-$0.4 &$-$0.15 & $-$0.129 & \quad 276184 & $-$0.071\,\, & $-$196\\
        & $3s3d$ & ${^3}D_1$ &  279758 &  279851 &  279830 &  279787 &  $-$221 \quad \quad &   0.0 &$-$0.03 & $-$0.010 & \quad 279566 &    0.069\,\, &    192\\
        & $3s3d$ & ${^3}D_2$ &  279773 &  279877 &  279860 &  279817 &  $-$230 \quad \quad &   0.0 &$-$0.03 & $-$0.016 & \quad 279587 &    0.067\,\, &    186\\
        & $3s3d$ & ${^3}D_3$ &  279804 &  279916 &  279906 &  279863 &  $-$245 \quad \quad &   0.0 &$-$0.04 & $-$0.021 & \quad 279618 &    0.066\,\, &    186\\
        & $3s3d$ & ${^1}D_2$ &  320925 &  322706 &  321257 &  321147 &  $-$222 \quad \quad &$-$0.6 &$-$0.10 & $-$0.069 & \quad 320925 &    0.000\,\, &    0 \\
        & $3p3d$ & ${^3}F_2$ &  383627 &  382362 &  383715 &  383722 &  $-$258 \quad \quad &   0.3 &$-$0.02 & $-$0.025 & \quad 383464 &    0.042\,\, &    163\\
        & $3p3d$ & ${^3}F_3$ &  384280 &  383037 &  384412 &  384420 &  $-$296 \quad \quad &   0.3 &$-$0.03 & $-$0.036 & \quad 384124 &    0.041\,\, &    156\\
        & $3p3d$ & ${^3}F_4$ &  385115 &  383896 &  385294 &  385302 &  $-$342 \quad \quad &   0.3 &$-$0.05 & $-$0.049 & \quad 384960 &    0.040\,\, &    155\\
        & $3p3d$ & ${^1}D_2$ &  389464 &  388983 &  389717 &  389687 &  $-$312 \quad \quad &   0.1 &$-$0.06 & $-$0.057 & \quad 389376 &    0.023\,\, &    88 \\
        & $3s4s$ & ${^3}S_1$ &  407323 &  405211 &  407496 &  407531 &  $-$180 \quad \quad &   0.5 &$-$0.04 & $-$0.051 & \quad 407351 & $-$0.007\,\, & $-$28 \\
        & $3p3d$ & ${^3}P_2$ &  409004 &  409106 &  409196 &  409142 &  $-$297 \quad \quad &   0.0 &$-$0.05 & $-$0.034 & \quad 408845 &    0.039\,\, &    159\\
        & $3p3d$ & ${^3}P_1$ &  409551 &  409657 &  409750 &  409696 &  $-$304 \quad \quad &   0.0 &$-$0.05 & $-$0.035 & \quad 409392 &    0.039\,\, &    159\\
        & $3p3d$ & ${^3}P_0$ &  409949 &  410028 &  410153 &  410101 &  $-$319 \quad \quad &   0.0 &$-$0.05 & $-$0.037 & \quad 409783 &    0.041\,\, &    166\\

\hline \hline
\end{tabular}
\end{table*}

\section{Results and Discussions}

In Tables \ref{tbl1:energies}~--~\ref{tbl3:energies} 
we compare with the experiment the results of our CI, CI+MBPT, and CI+AO ab initio calculations of the magnesium isoelectronic sequence. \tref{tbl1:energies} includes ions Mg I, Al II; \tref{tbl2:energies} and \tref{tbl3:energies} present results for Si III and P IV and for S V and Cl VI respectively. For the ground states we give the two-electron binding energies. These energies are equal to the sum of the first two ionization potentials. For all other states we give transition energies from the ground states. All energy values are in \cm. The same notations are used in all tables. To illustrate the accuracy of each of the theoretical approaches the relative differences of our results with the experiment are given in the columns designated as ``Diff.~with~expt.''.

\begin{figure*}[tbh]
\includegraphics[height=5.9cm, width=8.91cm]{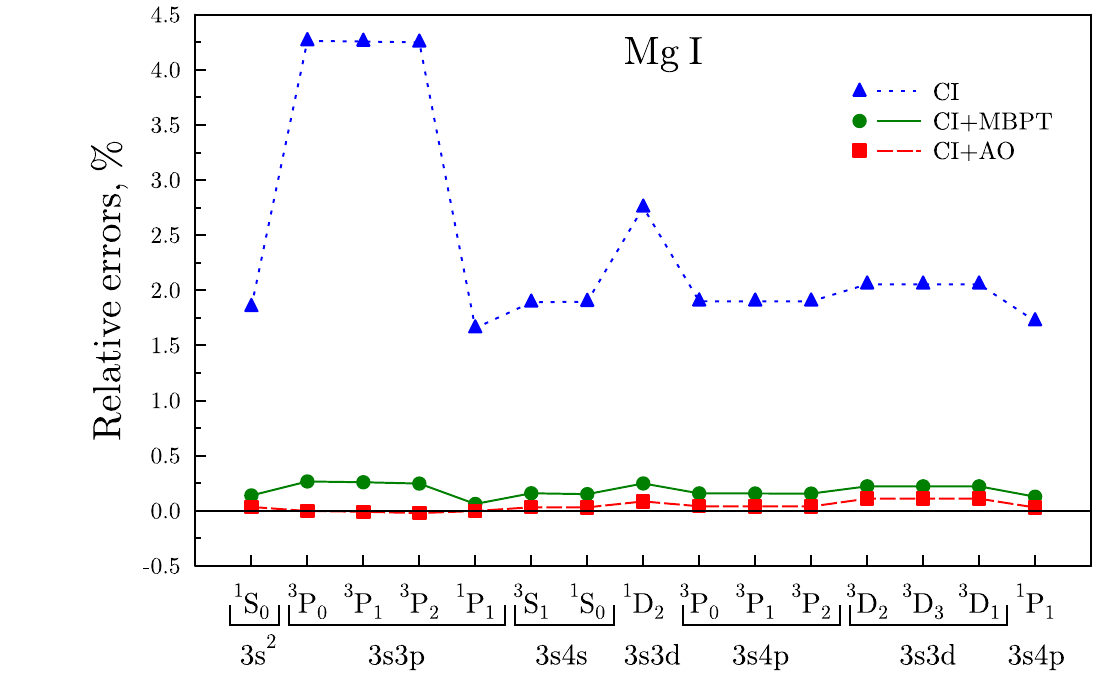}
\includegraphics[height=5.9cm, width=8.91cm]{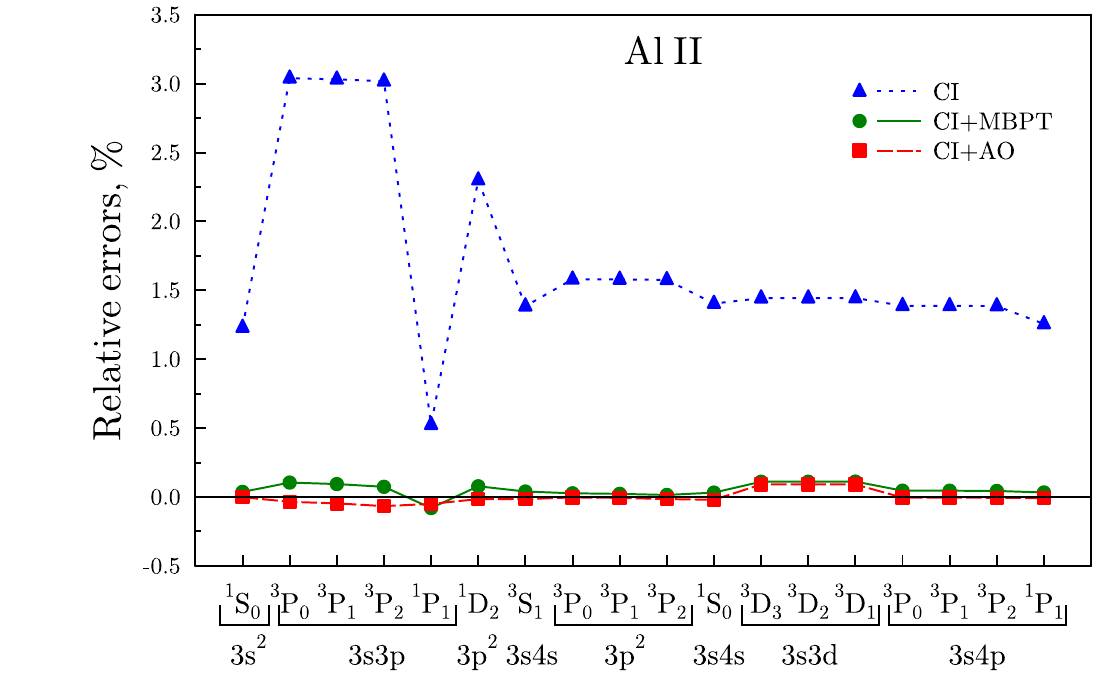}
\includegraphics[height=5.9cm, width=8.91cm]{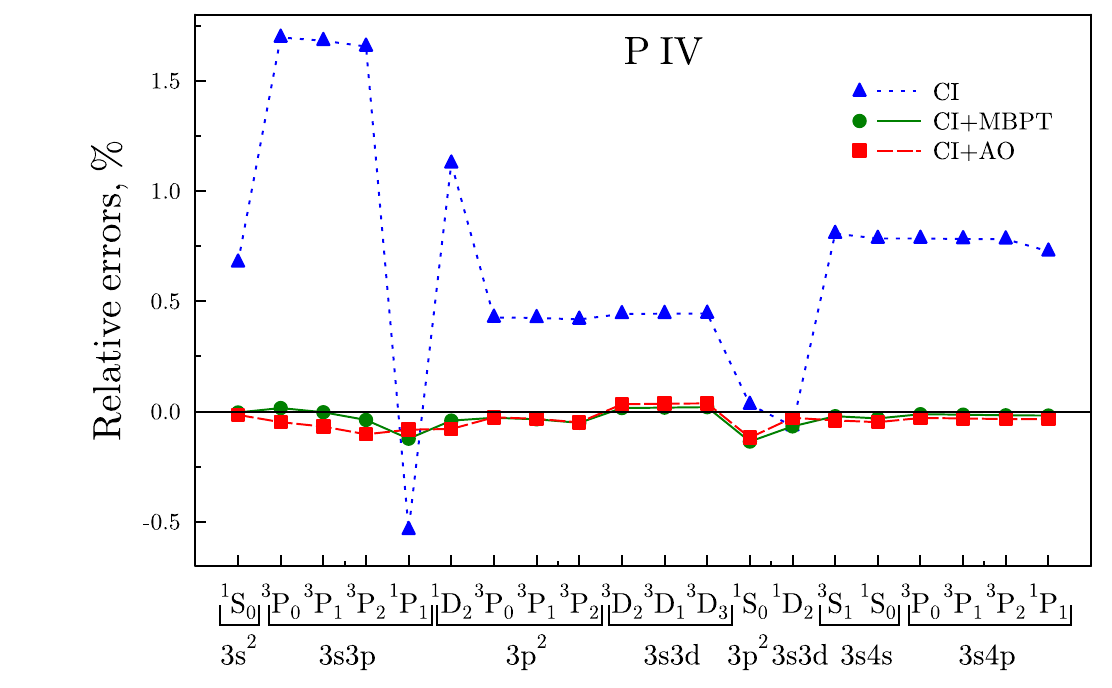}
\includegraphics[height=5.9cm, width=8.91cm]{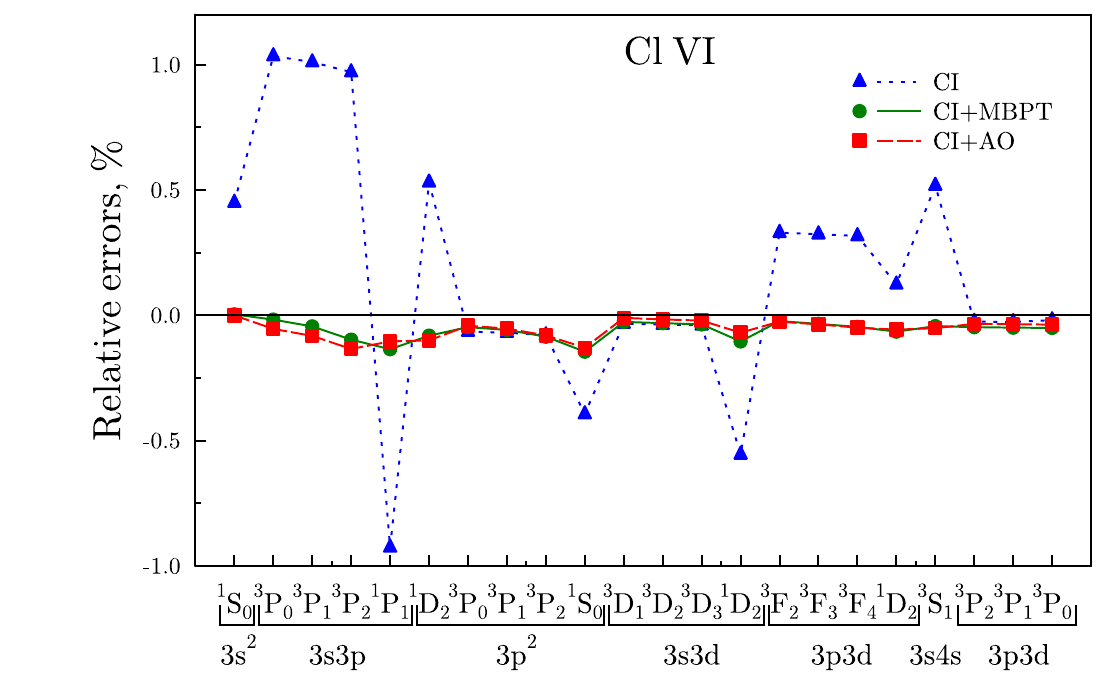}
\caption{The relative theoretical errors for Mg-like ions within CI, CI+MBPT, and CI+AO methods for the Dirac-Coulomb Hamiltonian.}
 \label{graphs_1}
\end{figure*}

\begin{figure*}[tbh]
\includegraphics[height=6.0cm, width=8.91cm]{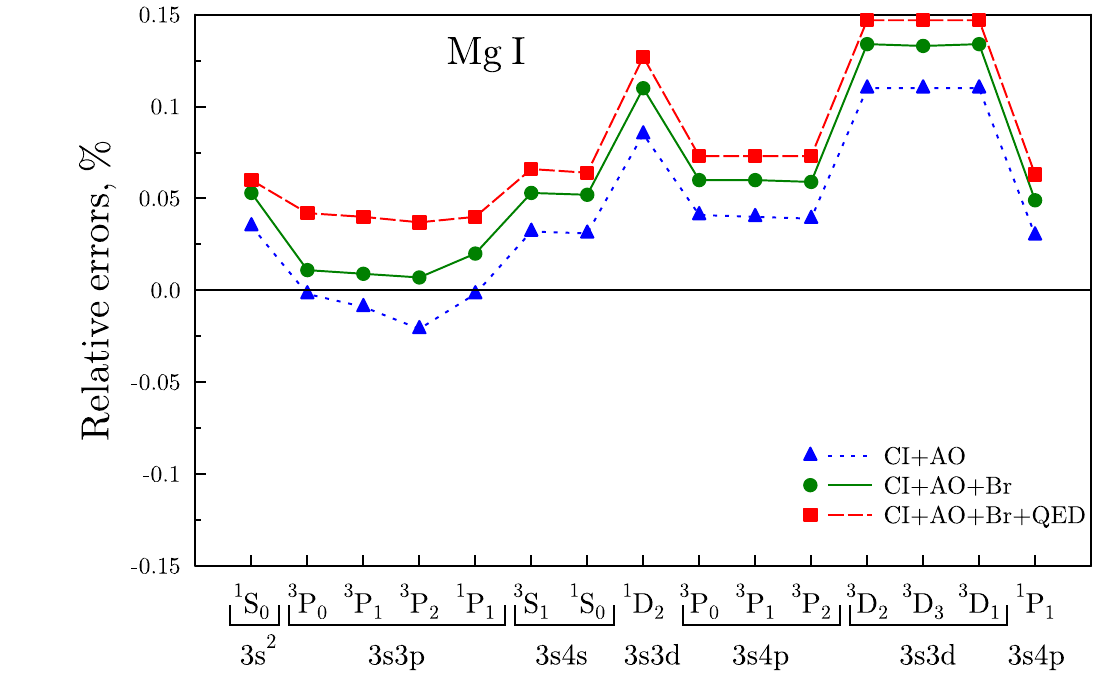}
\includegraphics[height=6.0cm, width=8.91cm]{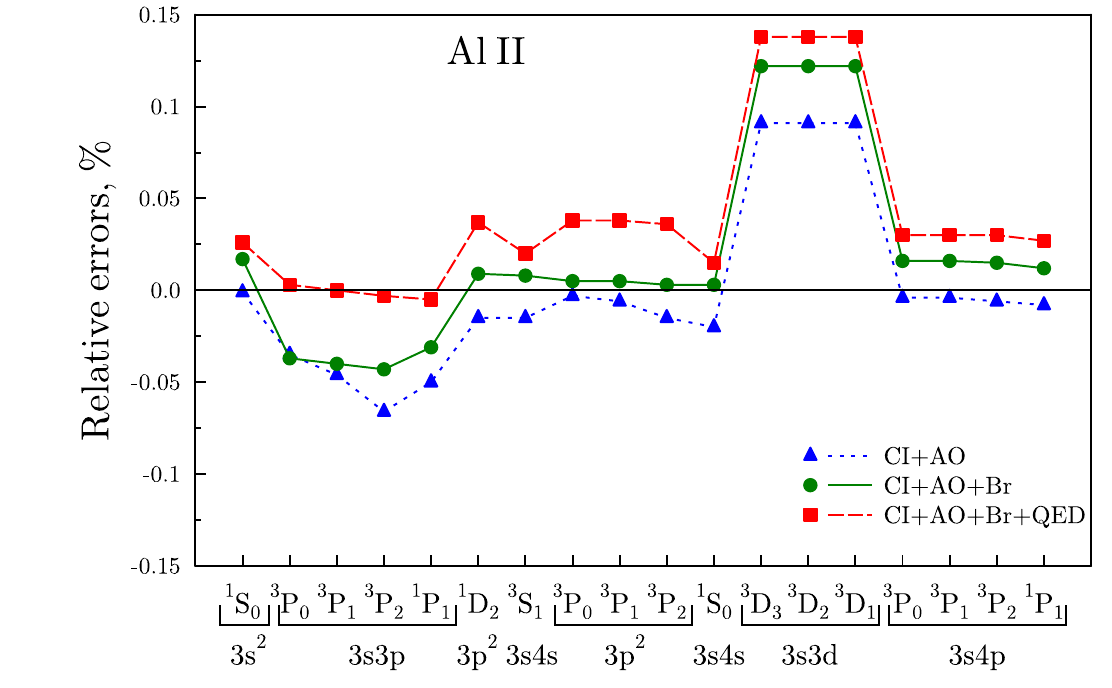}
\includegraphics[height=6.0cm, width=8.91cm]{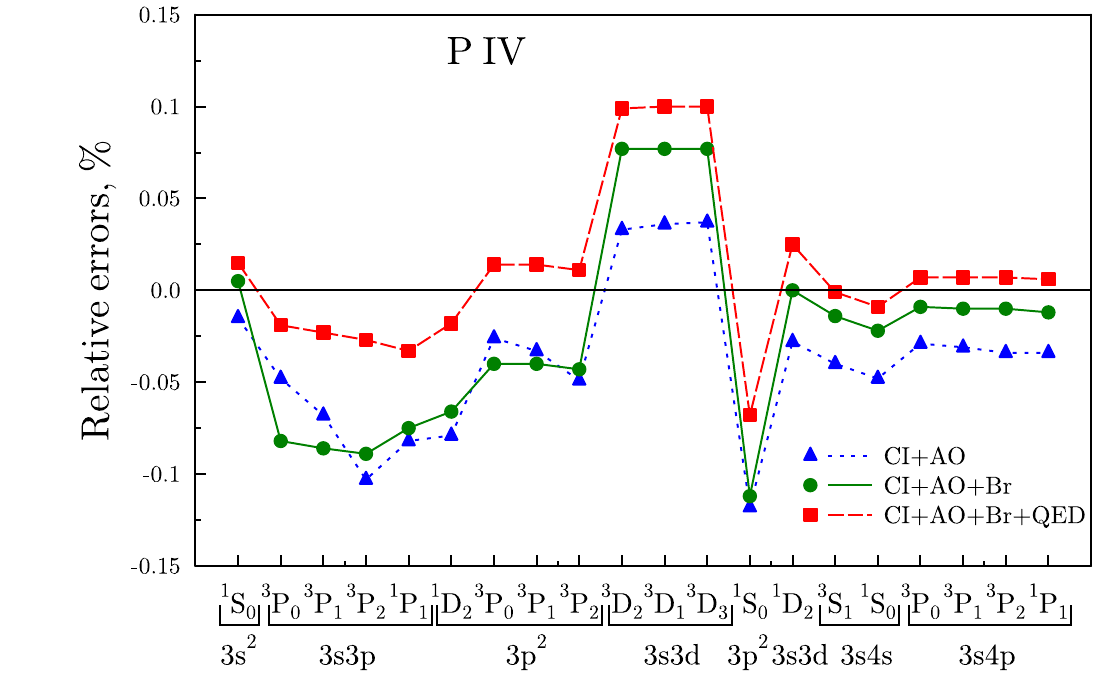}
\includegraphics[height=6.0cm, width=8.91cm]{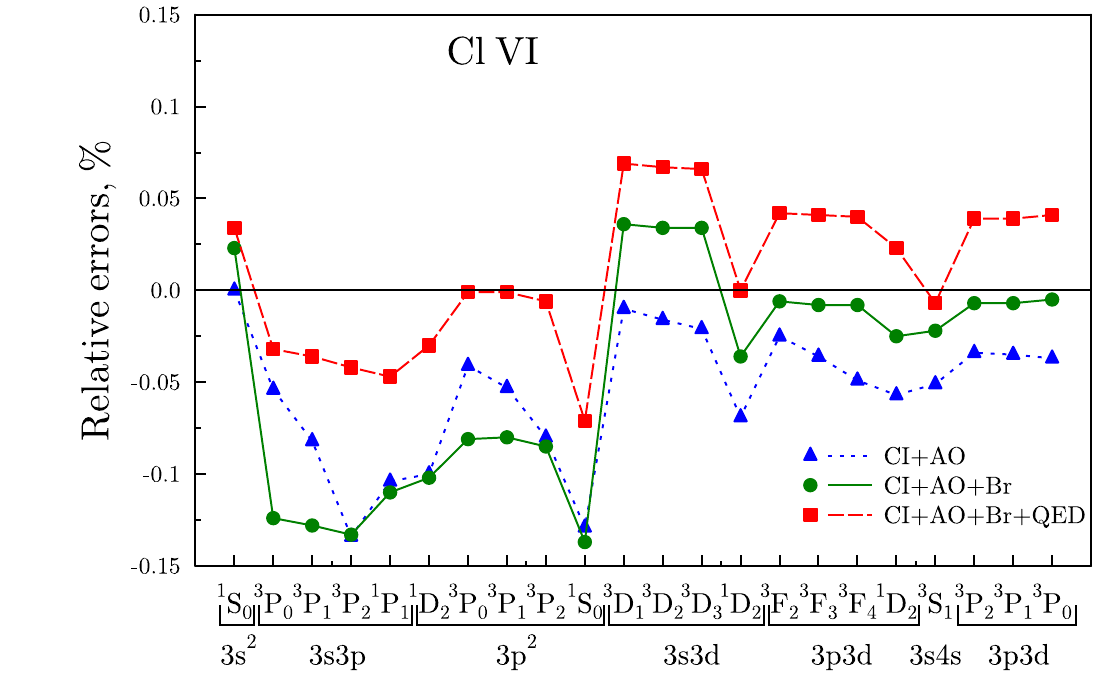}
\caption{Calculated relative errors for Mg-like ions within CI+AO method for Dirac-Coulomb Hamiltonian (triangles), Dirac-Coulomb-Breit Hamiltonian (circles), and Dirac-Coulomb-Breit Hamiltonian with QED corrections \eqref{LS7} (squares).}
 \label{graphs_2}
\end{figure*}

Our main concern here is the analysis of the accuracy of the three theoretical methods and the role of different corrections. On \fref{graphs_1} and \fref{graphs_2} we present relative theoretical errors for different methods for the four typical ions of the sequence including lightest and heaviest ones. For the ground states we again give the errors for the two-electron binding energies. For other states the errors correspond to the transition frequencies from the ground states. The Plots in \fref{graphs_1} demonstrate the accuracy of all three methods for the Dirac-Coulomb Hamiltonian. We see that for each method there is certain improvement along the isoelectronic sequence. For example, the average accuracy of the CI method for Mg I is about 2\%, and improves to roughly 0.4\% for the Cl VI. Similarly the accuracy of the CI+MBPT method improves from 0.2\% for Mg I to 0.1\% for Cl VI. At the same time the difference between CI+MBPT and CI+AO decreases with the ion charge $Z$, and almost disappears for Cl VI. This indicates smaller role of the higher order core-valence correlations for heavier ions. It is interested to note that the CI space for the triplet states was already saturated on the $15spdfg$ level. But in order to obtain similar accuracy for singlet states we had to increase the CI space to $21spdfg$.

\fref{graphs_2} demonstrates the influence of the Breit and QED corrections. These corrections are absolutely negligible for the CI calculations and only marginally noticeable for the CI+MBPT method. Therefore, \fref{graphs_2} presents results only for the CI+AO method. We see that for the light ions (Mg I, Al II) these corrections are negligible at the existing level of accuracy of treating electron correlations. However, for the heaviest ion (Cl VI) these corrections become essential and somewhat improve the final accuracy: an average error decreases from 0.06\% to 0.04\%. Note that for the ions considered here Breit and QED corrections are comparable. Breit interaction generally improves the fine structure splittings, while QED corrections decrease the overall scatter of errors.

Finally, we consider the energy dependence of the effective Hamiltonian. We find out that respective corrections are comparable in size to the difference between the CI+MBPT and CI+AO methods. It agrees with the conclusion in Ref.\ \cite{Koz03} that accurate treatment of the high-order corrections requires also including corrections on the energy dependence.  The average size of these corrections to the valence energies monotonously decrease from 0.06\% for Mg I to 0.02\% for Cl VI. In general they do not improve the agreement with the experiment. Only for the CI+MBPT calculations of Mg I some improvement (about 0.1\%) take place.
This may mean that corrections on the energy dependence cancel some high-order terms which are missing in our calculations. We conclude that for the present variant of the CI+AO method, which is based on the linearized SD CC, these corrections should be neglected. They also should not be included in the CI+MBPT calculations. Note that this significantly simplifies calculations with the package \cite{KPST15}.

\section{Conclusion}

In this paper we studied the accuracy of the CI+AO method \cite{Koz04,SKJJ09} for the isoelectronic sequence of Mg. These ions have ten electrons in the closed shells and two valence electrons and are often used as a test ground for the atomic theory. We found out that CI+AO method provides higher accuracy than the simpler and more common CI+MBPT method \cite{DFK96b}. While the accuracy of the CI+MBPT method was on the level 0.1~--~0.2\%, the accuracy of the CI+AO was roughly two times higher, 0.05~--~0.1\%. Note that conventional valence CI is an order of magnitude less accurate.

The accuracy slightly increases along the isoelectronic sequence. For the first member of the sequence, Mg I, the final accuracy of the theory for the low-lying levels is close to 0.08\%  and for the last ion, Cl VI, it is about 0.04\%.  Breit and QED corrections start to become important on this level of accuracy for the atoms and ions with $Z\gtrsim 20$. Retardation part of the Breit interaction is known to be significantly smaller than magnetic part and can be still neglected. For QED corrections it is sufficient to account only for the $s$-wave contribution and use simplified semiempirical expression \eqref{LS7}.

We also studied corrections on the energy dependence of the effective Hamiltonian in the CI+AO method. On the one hand, we found them to be rather small. On the other hand, these corrections did not improve agreement with the experiment. We conclude that corrections on the energy dependence can be neglected for the present variant of the CI+AO method, when the all-order part corresponds to the linearized coupled cluster method in the SD approximation. This significantly simplifies calculations and makes the whole method more practical.

\acknowledgments
We are grateful to S.~Porsev, M.~Safronova, and I.~Tupitsyn for useful discussions. 
This work is partly supported by the Russian Foundation for Basic Research Grant No.~14-02-00241.


\end{document}